\documentclass[twocolumn]{aastex631}

\newcommand\LCDM{\Lambda\mathrm{CDM}}
\newcommand\Morpheus{\emph{Morpheus}}
\newcommand\WebbPSF{\emph{WebbPSF}}
\newcommand\ProFit{\emph{ProFit}}
\newcommand\ForcePho{\emph{forcepho}}
\newcommand\JWST{\emph{JWST}}

\newcommand\HST{\emph{HST}}
\newcommand\ALMA{\emph{ALMA}}
\newcommand\photutils{\emph{photutils}}
\newcommand\spheroid{\emph{spheroid}}
\newcommand\disk{\emph{disk}}
\newcommand\irregular{\emph{irregular}}
\newcommand\irr{\emph{irregular}}
\newcommand\psc{\emph{compact/point source}}
\newcommand\background{\emph{background}}
\newcommand\av[1]{\langle#1\rangle}

\shorttitle{\JWST{} Discovery of High-Redshift Disk Galaxy Candidates}
\shortauthors{Robertson et al.}
\graphicspath{{./}{figures/}}

\begin{document}

\title{\emph{Morpheus} Reveals Distant Disk Galaxy Morphologies with \emph{JWST}: \\The First AI/ML Analysis of \emph{JWST} Images}

\author[0000-0002-4271-0364]{Brant E. Robertson}
\affiliation{Department of Astronomy and Astrophysics, University of California, Santa Cruz, 1156 High Street, Santa Cruz, CA 95064, USA}

\author[0000-0002-8224-4505]{Sandro Tacchella}
\affiliation{Kavli Institute for Cosmology, University of Cambridge, Madingley Road, Cambridge, CB3 0HA, UK}
\affiliation{Cavendish Laboratory, University of Cambridge, 19 JJ Thomson Avenue, Cambridge, CB3 0HE, UK}

\author[0000-0002-9280-7594]{Benjamin D. Johnson}
\affiliation{Center for Astrophysics $\vert$ Harvard \& Smithsonian, 60 Garden Street, Cambridge, MA 02138, USA}

\author[0000-0002-8543-761X]{Ryan Hausen}
\affiliation{Department of Physics and Astronomy, The Johns Hopkins University, 3400 N. Charles St., Baltimore, MD 21218, USA}

\author[0000-0002-3475-587X]{Adebusola B. Alabi}
\affiliation{Department of Astronomy and Astrophysics, University of California, Santa Cruz, 1156 High Street, Santa Cruz, CA 95064, USA}

\author[0000-0003-4109-304X]{Kristan ~Boyett}
\affiliation{School of Physics, University of Melbourne, Parkville 3010, VIC, Australia}
\affiliation{ARC Centre of Excellence for All Sky Astrophysics in 3 Dimensions (ASTRO 3D), Australia}

\author{Andrew J. Bunker}
\affiliation{Department of Physics, University of Oxford, Denys Wilkinson Building, Keble Road, Oxford OX13RH, U.K.}

\author[0000-0002-6719-380X]{Stefano Carniani}
\affiliation{Scuola Normale Superiore, Piazza dei Cavalieri 7, I-56126 Pisa, Italy}

\author[0000-0003-1344-9475]{Eiichi Egami}
\affiliation{Steward Observatory, University of Arizona, 933 N. Cherry Avenue, Tucson, AZ 85721, USA}

\author[0000-0002-2929-3121]{Daniel J. Eisenstein}
\affiliation{Center for Astrophysics $\vert$ Harvard \& Smithsonian, 60 Garden Street, Cambridge, MA 02138, USA}

\author[0000-0001-9262-9997]{Kevin N. Hainline}
\affiliation{Steward Observatory, University of Arizona, 933 N. Cherry Avenue, Tucson, AZ 85721, USA}

\author[0000-0003-4337-6211]{Jakob M. Helton}
\affiliation{Steward Observatory, University of Arizona, 933 N. Cherry Avenue, Tucson, AZ 85721, USA}

\author[0000-0001-7673-2257]{Zhiyuan Ji}
\affiliation{University of Massachusetts Amherst, 710 North Pleasant Street, Amherst, MA 01003-9305, USA}

\author[0000-0002-5320-2568]{Nimisha Kumari}
\affiliation{AURA for the European Space Agency, Space Telescope Science Institute, 3700 San Martin Drive, Baltimore, MD 21218, USA}

\author[0000-0002-6221-1829]{Jianwei Lyu}
\affiliation{Steward Observatory, University of Arizona, 933 N. Cherry Avenue, Tucson, AZ 85721, USA}

\author[0000-0002-4985-3819]{Roberto Maiolino}
\affiliation{Kavli Institute for Cosmology, University of Cambridge, Madingley Road, Cambridge, CB3 0HA, UK}
\affiliation{Cavendish Laboratory, University of Cambridge, 19 JJ Thomson Avenue, Cambridge, CB3 0HE, UK}

\author[0000-0002-7524-374X]{Erica J. Nelson}
\affiliation{Department for Astrophysical and Planetary Science, University of Colorado, Boulder, CO 80309, USA}

\author[0000-0002-7893-6170]{Marcia J. Rieke}
\affiliation{Steward Observatory, University of Arizona, 933 N. Cherry Avenue, Tucson, AZ 85721, USA}

\author[0000-0003-4702-7561]{Irene Shivaei}
\affiliation{Steward Observatory, University of Arizona, 933 N. Cherry Avenue, Tucson, AZ 85721, USA}

\author[0000-0002-4622-6617]{Fengwu Sun}
\affiliation{Steward Observatory, University of Arizona, 933 N. Cherry Avenue, Tucson, AZ 85721, USA}

\author[0000-0003-4891-0794]{Hannah \"Ubler}
\affiliation{Kavli Institute for Cosmology, University of Cambridge, Madingley Road, Cambridge, CB3 0HA, UK}
\affiliation{Cavendish Laboratory, University of Cambridge, 19 JJ Thomson Avenue, Cambridge, CB3 0HE, UK}

\author[0000-0003-2919-7495]{Christina C. Williams}\affiliation{NSF's National Optical-Infrared Astronomy Research Laboratory, 950 N. Cherry Avenue, Tucson, AZ 85719, USA}\affiliation{Steward Observatory, University of Arizona, 933 N. Cherry Avenue, Tucson, AZ 85721, USA} 

\author[0000-0001-9262-9997]{Christopher N. A. Willmer}
\affiliation{Steward Observatory, University of Arizona, 933 North  Cherry Avenue,Tucson, AZ 85721, USA}

\author[0000-0002-7595-121X]{Joris Witstok}
\affiliation{Kavli Institute for Cosmology, University of Cambridge, Madingley Road, Cambridge, CB3 0HA, UK}
\affiliation{Cavendish Laboratory, University of Cambridge, 19 JJ Thomson Avenue, Cambridge, CB3 0HE, UK}

\begin{abstract}
The dramatic first images with \emph{James Webb Space Telescope} (\JWST{})
demonstrated its power to provide unprecedented spatial detail
for galaxies in the high-redshift universe. Here, we leverage the
resolution and depth of the \JWST{} Cosmic Evolution Early Release
Science Survey (CEERS) data
in the Extended Groth Strip (EGS) to perform
pixel-level morphological classifications of galaxies in
\JWST{} $F150W$ imaging using the \Morpheus{} deep learning 
framework for astronomical image analysis. By cross-referencing with
existing photometric redshift catalogs from the \emph{Hubble Space Telescope} (\HST{})
CANDELS survey, we show that
\JWST{} images indicate the emergence of \emph{disk} morphologies before $z\sim2$
and with candidates appearing as early as $z\sim5$. By modeling the light profile of
each object and accounting for the \JWST{} point-spread function, we find
the high-redshift disk candidates have exponential surface brightness profiles with
an average \citet{sersic1968a} index $\av{n}=1.04$ and $>90\%$ displaying ``disky'' profiles ($n<2$).
Comparing with 
prior \Morpheus{} classifications in CANDELS
we find that a plurality of \JWST{} disk galaxy candidates were previously 
classified as \emph{compact} based on the shallower \HST{} imagery, indicating
that the improved optical quality and depth of the \JWST{}
helps to reveal disk morphologies that were hiding in the noise.
We discuss the implications of these early
disk candidates on theories for cosmological disk galaxy formation.
\end{abstract}

\keywords{}

\section{Introduction}
\label{sec:intro}

The formation and evolution
of disk galaxies in the context
of our $\LCDM$ cosmology remains
as one of the most complex problems
in astrophysics. The violent,
hierarchical nature of galaxy
mass assembly at high redshifts
can destroy dynamically
cold disks, and yet thin disk galaxies
are plentiful in the local universe.
Determining the era when the first
disk galaxies could form would provide
an important milestone in understanding how
galaxy morphology develops over cosmic
time. This \emph{Letter} presents first
results of applying artificial intelligence/machine learning (AI/ML) techniques to 
\emph{James Webb Space Telescope} (\JWST) images,
searching for distant disk galaxies by
analyzing \JWST{} data with the 
\Morpheus{} deep learning framework
\citep[][hereafter H20]{hausen2020a}.

Observations of disk galaxy formation in the
distant universe have advanced greatly over
the last two decades.
Integral field spectroscopy of 
disk galaxies at intermediate redshifts ($z\sim2$)
established that the ionized gas dynamics of
many early-forming disks differed substantially from
those in the local universe, with lower ratios
of rotational velocity to velocity dispersion $V/\sigma$ than in the present day \citep[e.g.,][]{genzel2006a,law2007a,genzel2008a,stott2016a,simons2017a,forster-schreiber2018a,wisnioski2019a}.
Large, star-forming disk galaxies at redshifts
$z\sim1-3$ can be baryon-dominated, with relatively
little dark matter contributing to their rotation curves
\citep{wuyts2016a,genzel2017a,ubler2018a,price2020a,genzel2020a}.
The structure of early disks can be
distinct from dynamically colder disk galaxies
at the present, perhaps reflecting a transition between
disks heavily influenced by the hierarchical mass assembly
process to more ordered rotation at lower
redshifts \citep[e.g.,][]{kassin2012a,simons2017a,tiley2021a}.
This transition may mirror the evolution of observed axis ratios, as
observations with \HST{} have established a smaller fraction of
disk galaxies at earlier cosmic times \citep[e.g.,][]{law2012a,vanderwel2014b,zhang2019a}.

Observations of [CIII] and CO emission at even higher redshifts
with the \emph{Atacama Large Millimeter/Submillimeter Array}
(\ALMA)
have discovered evidence for rotationally-supported
disk galaxies in the first two billion years
of cosmic history
\citep{smit2018a,neeleman2020a,rizzo2020a,rizzo2021a,fraternali2021a,lelli2021a,tsukui2021a}.
These $z>4$
galaxies show $V/\sigma\sim10-20$,
substantially larger than those predicted for the
bulk population of disk galaxies
by cosmological simulations \citep[e.g.,][]{pillepich2019a}.
Recent theoretical work suggests that relatively
massive galaxies may sustain cold disks even in the
early universe \citep{kretschmer2022a,gurvich2022a},
but understanding how rare such objects are, elucidating the role of the kinematical tracer, and determining their relationship to the broader galaxy population
all require more investigation.

\begin{figure*}[ht!]
\epsscale{1.15}
\plotone{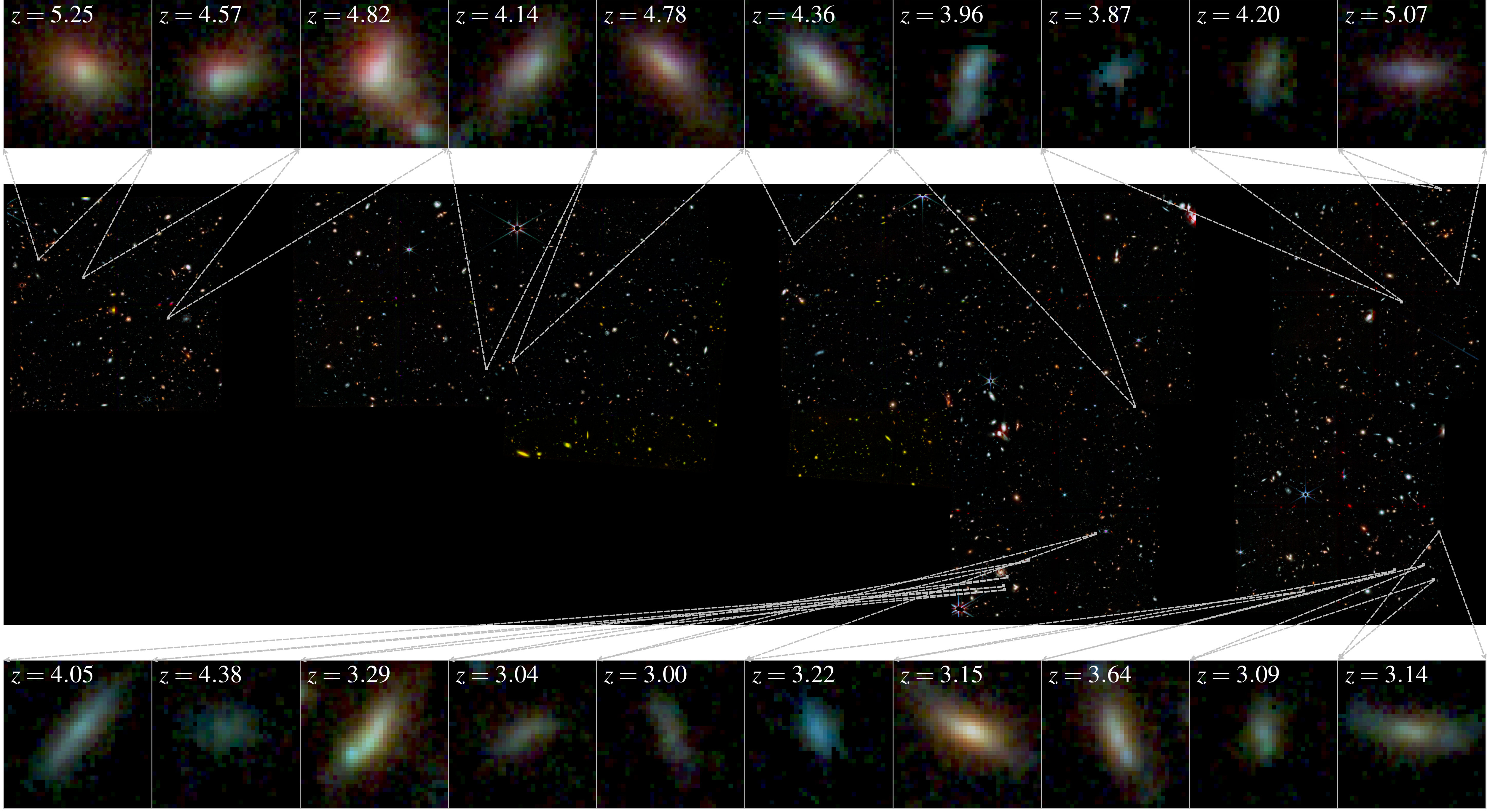}
\caption{High-redshift disk galaxy
candidates selected by the \Morpheus{}
AI/ML classifier
in the CEERS EGS \JWST{}
imagery. Shown is a \JWST{} $F444W$-$F200W$-$F115W$
RGB false-color image (center) along with
thumbnails of the 20 highest redshift disk galaxy
candidates (top and bottom rows). Most disk
galaxy candidates selected by \Morpheus{}
show flattened, extended
light distributions consistent with disk
morphologies.
\label{fig:composite}}
\end{figure*}

The new \JWST{} data can address outstanding questions raised by these previous results.
When did the first disk galaxy candidates appear
in the universe? What is the relative gas to 
stellar mass content of these galaxies, and how
does that influence their formation? \JWST{} can
contribute substantially to answering these questions,
both through better constraints on the stellar
population properties of early disks and through
kinematic measures via its unprecedented spectroscopic
capabilities.
Addressing these
questions with \JWST{} will require the
identification of more
high-redshift disk galaxy candidates. Here, 
we perform pixel-level analysis of the 
\JWST{} Cosmic Evolution Early Release
Science Survey (CEERS; Finkelstein et al., in prep.) data
in the Extended Groth Strip (EGS) using 
the \Morpheus{} deep learning model (H20)
to search for high-redshift ($z>2$) disk
galaxy candidates. \Morpheus{}
has been trained to classify 
galaxies, stars, and sky pixels in 
space-based astronomical images, and
we apply the model directly to the
EGS \JWST{} $F150W$ mosaic without modification.
We identify galaxies with photometric
redshifts $z>2$ with dominant disk
morphologies as determined by \Morpheus{}.
We then verify via surface brightness
model fitting that the vast majority of
candidates are structurally consistent
with being high-redshift disks.
These results indicate that AI/ML methods
like \Morpheus{} will effectively identify
distant disk galaxy candidates for kinematic
follow-up observations with spectroscopy.
Our AI/ML approach complements other
recent work identifying disk galaxy candidates
in deep \JWST{} images through traditional analyses 
\citep[e.g.,][]{fudamoto2022a,nelson2022a,ferreira2022a,jacobs2022a,wu2022a}.
Where necessary, we assume parameters
for the flat $\Lambda$CDM cosmology
determined by the
\citet{planck2020a} survey (i.e., $h=0.674$, $\Omega_m=0.315$).

\section{Method}
\label{sec:method}

The identification of high-redshift disk
galaxy candidates in the \JWST{} CEERS images
proceeded by first fully re-reducing 
the imaging data products and
then performing object detection
and photometry (\S \ref{sec:images}).
We then analyzed the \JWST{} $F150W$
images with \Morpheus{} to identify
objects with disk morphologies
(\S \ref{sec:morph}), and
cross-referenced with catalogs
from the literature to select disk
candidates at $z>2$ (\S \ref{sec:selection}).

\subsection{Imaging, Detection, and Photometry}
\label{sec:images}

The \JWST{} CEERS EGS images \citep[34.5 arcmin$^{2}$, see ][ Finkelstein in prep.]{finkelstein2022a}
were downloaded and
processed with the STScI \JWST{}
pipeline ({\tt v1.6.2}) using custom treatments for data artifacts to remove ``wisps'' and ``snowballs'' \citep[e.g., \S 6 of ][]{rigby2022a},
background subtraction and sky flats ({\tt jwst\_0953.pmap}), and $1/f$-noise correction.
The exposures were astrometrically aligned to the CEERS \HST{} F160W image
and mosaiced onto a 0.03''/pixel image.

Object detection and photometry were 
performed using a set of custom scripts
based on \photutils{} methods.
We created a detection image from
an inverse-variance weighted
stack of the long-wavelength $F277W$,
$F356W$, $F410M$, and $F444W$ mosaics
to provide a deep, rest-frame optical selection.
Contiguous regions in the detection image with
pixels $SNR>5$ were identified as objects
in a segmentation map, and
forced aperture photometry was performed about their
centroids.

\subsection{Morphological Analysis}
\label{sec:morph}

The \JWST{} $F150W$ images were analyzed
using a version of \Morpheus{} (H20) trained on \HST{} F160W images at a similar observed wavelength.
\Morpheus{} is a UNet \citep{ronneberger2015a}
convolutional neural network with
residual links \citep{he2015a} that performs
pixel-level classification, assigning a 
``model probability'' that each pixel
in a science-quality FITS image \citep{wells1981a} 
belongs to \spheroid{}, \disk{}, \irregular{},
\psc{}, or \background{} classes. A ``classification
image'' for each class is produced to record
the pixel-level model probabilities, which can
be agglomerated into object-level classifications
by summations over the pixels.
H20
used the visual classifications of CANDELS
galaxies provided by \citet{kartaltepe2015a} (90\% completeness limit for an $H<24.5$AB selection)
to train
the model, and then used \Morpheus{} to classify
all the pixels in CANDELS by processing the \HST{}
$F606W$, $F814W$, $F125W$, and $F160W$ images. \Morpheus{} has
since been updated to include an
accelerated attention mechanism \citep{wang2017a, vaswani2017a, hu2020a}
that improves the model performance by using
non-local image features
and allows for larger regions of the image to be
classified at once, providing a roughly 100$\times$
speed up in classification efficiency relative to
H20 with comparable performance \citep[for a discussion of related changes to \Morpheus{}, see][]{hausen2022a}. The
updated \Morpheus{} model was previously
trained to perform
classification on \HST{} $F160W$ images alone using the Adam 
optimizer \citep{kingma2014a}. We leverage this training to provide a first attempt at classification
of the \JWST{} $F150W$ images at a similar observed wavelength and
validate our classifications with surface
brightness profile modeling.

\subsection{Disk Galaxy Candidate Selection}
\label{sec:selection}

To identify disk galaxy candidates, we used
the \Morpheus{} classification images with the
segmentation maps and detection images (see \S \ref{sec:images})
to create SNR-weighted object-level 
model probabilities for each class. We
cross-matched our sources against the 
CANDELS v1 EGS photometric redshift and
stellar population synthesis analysis
catalogs \citep{stefanon2017a}, and 
then selected all objects at
redshifts $z\ge2$ with a \Morpheus{}-assigned
object-level model probability $p(\disk{})>0.5$.
For this initial analysis, we restrict our
sample to objects with photometric redshift
estimates available in the CANDELS catalogs
\citep[90\% complete for point sources $F160W<26.62$, see][ for details on detection and photometric redshift determinations]{stefanon2017a} and spatially
close centroid matches ($\lesssim5$ pixels).
Over the NIRCam EGS footprint, we find
good spatial matches to $\sim7700$ sources
in the CANDELS catalog
before redshift or morphological cuts.
Over all redshifts, about 1600 CANDELS EGS
sources are classified as \disk{} by \Morpheus{}.

\section{Results}
\label{sec:results}

The selection criteria (\S \ref{sec:selection}) 
applied to the
analysis of \JWST{} NIRCam images with 
\Morpheus{} (\S \ref{sec:morph})
identified 202 disk galaxy candidates out of 2507 catalog matches
with redshifts $z\ge2$.
The average photometric redshift of the sample
is $\av{z}=2.67$, with 42 disk candidates
at redshifts $z>3$ (out of 988) and 10 disk candidates
at redshifts $z>4$ (out of 353).
Below, we
examine the visual morphologies of
the objects (\S \ref{sec:visual})
and perform surface brightness model
fitting to verify consistency
with structural disk galaxy properties
(\S \ref{sec:sersic}).

\subsection{Visual Morphologies}
\label{sec:visual}

Figure
\ref{fig:composite} shows a 
\JWST{} $F444W$-$F200W$-$F115W$ RGB
false color image of the CEERS
EGS mosaic, highlighting 20 candidates
with large
CANDELS photometric redshifts
as 0.3''$\times$0.3'' thumbnails above and below
the mosaic. Many candidates
show disk-like visual morphologies,
often with flattened, extended
light distributions. Some
galaxies show central concentrations,
clumps, or amorphous light distributions
in addition to their extended disk morphologies.
Of the 202 high-redshift disk candidates, only 21\% were brighter
than the $H<24.5$AB completeness limit
for visual classification by \citet{kartaltepe2015a}
and of these 57\% were classified as \disk{} in our
prior \HST{}-based analysis.
Of the 160 objects fainter than $H=24.5$AB, only
8 (5\%) were previously classified as \disk{}.
The plurality (49\%) of objects
were previously classified as \psc{} as the 
\HST{} imaging quality was not detailed or deep enough
for the model to choose a distinctive morphology
for many objects. The remaining objects were previously classified
as \spheroid{} (5\%), \irr{} (16\%), or not detected by
the algorithm and labeled as \background{} (25\%).
To compare with previous AI/ML classifications, over
800 arcmin$^{2}$ of \HST{} CANDELS images \citet{huertas-company2016a}
reported finding 413 objects with
redshifts $z>2$
that they classify as disks, whereas with \Morpheus{} applied to 
\JWST{} $F150W$ we find 202 candidate
\disk{} galaxies at $z>2$ over only 34.5 arcmin$^{2}$.
Given the
wide range of redshifts $z\sim2-5$,
the colors of the objects vary
with $F115W-F150W\approx0-1$AB and
$F150W-F200W\approx-0.1-0.6$AB,
and mean colors $\av{F115W-F150W}=0.51$AB
and $\av{F150W-F200W}=0.17$AB.
From the CANDELS catalogs, we find that the
typical stellar masses are $\av{\log M_{\star}/M_\sun} = 9.13\pm0.48$,
with 40\% of objects inferred to have less than a billion
solar masses in stars.

\subsection{Surface Brightness Profiles}
\label{sec:sersic}

Following the identification of 
high-redshift disk galaxy 
candidates by \Morpheus{}, 
an evaluation of the
candidates through a traditional
morphological analysis can
help assess the efficacy of
our AI/ML approach even without
retraining on \JWST{} images.
If traditional metrics of
``diskiness'' agree with the
\Morpheus{} results, our
confidence in the AI/ML
classifications would be
strengthened.
In \citet{hausen2020a}, the
ability of \Morpheus{} to 
classify \citet{sersic1968a}
profiles based on traditional
associations between 
surface brightness profile
and morphology was demonstrated.
\Morpheus{} classifies extended
objects with $n=1$ S\'ersic
indices (exponential profiles) as
\disk{}, whereas objects with $n\ge4$
indices are usually classified as \spheroid{}
or \psc{}. 
In this work, we used the \ForcePho{}
(Johnson et al., in prep.) and \ProFit{}
\citep{robotham2017a} Bayesian profile
fitting codes to model the surface
brightness profile of each high-redshift
disk
galaxy candidate selected by \Morpheus{}.
We note that \Morpheus{} can spatially distinguish between the bulge and disk components of galaxies, and can also classify pixels in the star-forming regions of disks as \irr{} (see H20). We will explore these composite morphologies in \JWST{} imagery in future work.

\subsubsection{ForcePho Modeling}
\label{sec:forcepho}
When using \ForcePho{}, a
3'' $\times$ 3'' region about each
object centroid was cutout from the 
\JWST{} $F115$, $F150W$, and $F200W$
flux and flux error images. Source
locations from the detection catalog
were supplied to \ForcePho{}, which
then fit S\'ersic models
to the surface brightness profiles of
each object in the cutout, accounting
for possible flux contributions from
neighboring sources.
The PSFs used in convolving the models
to the data resolution were generated from a mixture
of Gaussians matched to reproduce the pre-flight
\WebbPSF{} 
\citep{perrin2014a} model for each \JWST{} filter.
The \ForcePho{} code performs Monte Carlo
Markov Chain sampling of the nine free
parameters including
$F115W$, $F150W$, $F200W$ fluxes,
the RA and DEC of each object, the isophotal
position angle, the axis ratio, the S\'ersic
index, and the half-light radius. This 
procedure allows for uncertainties on the
model parameters to be computed and possible
parameter covariances to be studied. A
single profile shape is fit to each filter by
allowing only the amplitude to vary between
them, and we do not attempt here to model
wavelength-dependent shape variations.

\subsubsection{ProFit Modeling}
\label{sec:profit}
For the \ProFit{} modeling, a
3'' $\times$ 3'' region about each
object centroid was cutout from the \JWST{} $F150W$
flux and flux error images. These images, along
with corresponding cutouts from the segmentation
map and a PSF model generated by \WebbPSF{}, were supplied to \ProFit{}
and a single-component S\'ersic model was fit to each image.
We used eight free parameters, including the
pixel centroids, total flux, effective radius,
S\'ersic profile indices ($0.5<n<20$), axis ratio, position angle, and
isophotal boxiness.
One hundred and ninety out of
202 objects received successful \ProFit{} fits, with the
reduced
$\av{\chi_\nu^2}=0.92\pm0.30$ for the sample. Failed \ProFit{} single component fits showed evidence for multiple components, and many of these objects were successfully modeled with \ForcePho{}.

\subsubsection{Surface Brightness Profile Model Results}
\label{sec:sbmod}

\begin{figure}[ht!]
\plotone{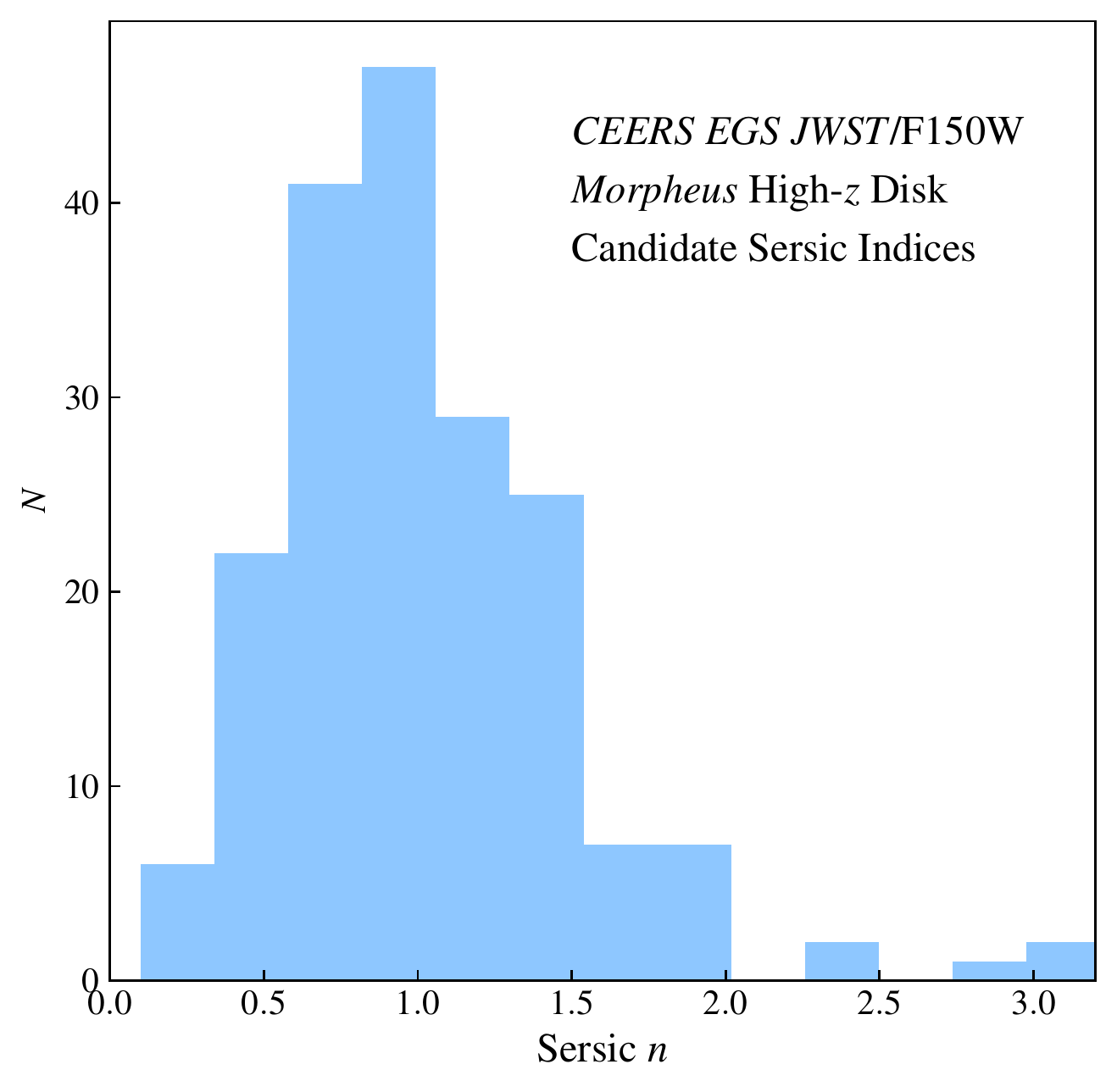}
\caption{High-redshift disk candidate
surface brightness profile
\citep{sersic1968a} indices determined by
\ProFit{} \citep{robotham2017a}. Shown
is a histogram of the best-fit S\'ersic index 
determined by fits to a 3'' $\times$ 3'' 
cutout of the \JWST{} $F150W$ flux and error
images about each candidate selected by \Morpheus{}.
The mean S\'ersic index of the sample is $\av{n}=1.04$,
characteristic of exponential profiles
traditionally associated
with disk galaxies. A single source with a S\'ersic index $n=4.9$ was omitted from the histogram for clarity.
Profile fits with \ForcePho{} (Johnson et al., in prep.) provide independent confirmation of the distribution, with 
97\% of high-redshift disk candidates displaying $n<2$.
\label{fig:sersic}}
\end{figure}

\begin{figure*}[ht!]
\epsscale{1.15}
\plotone{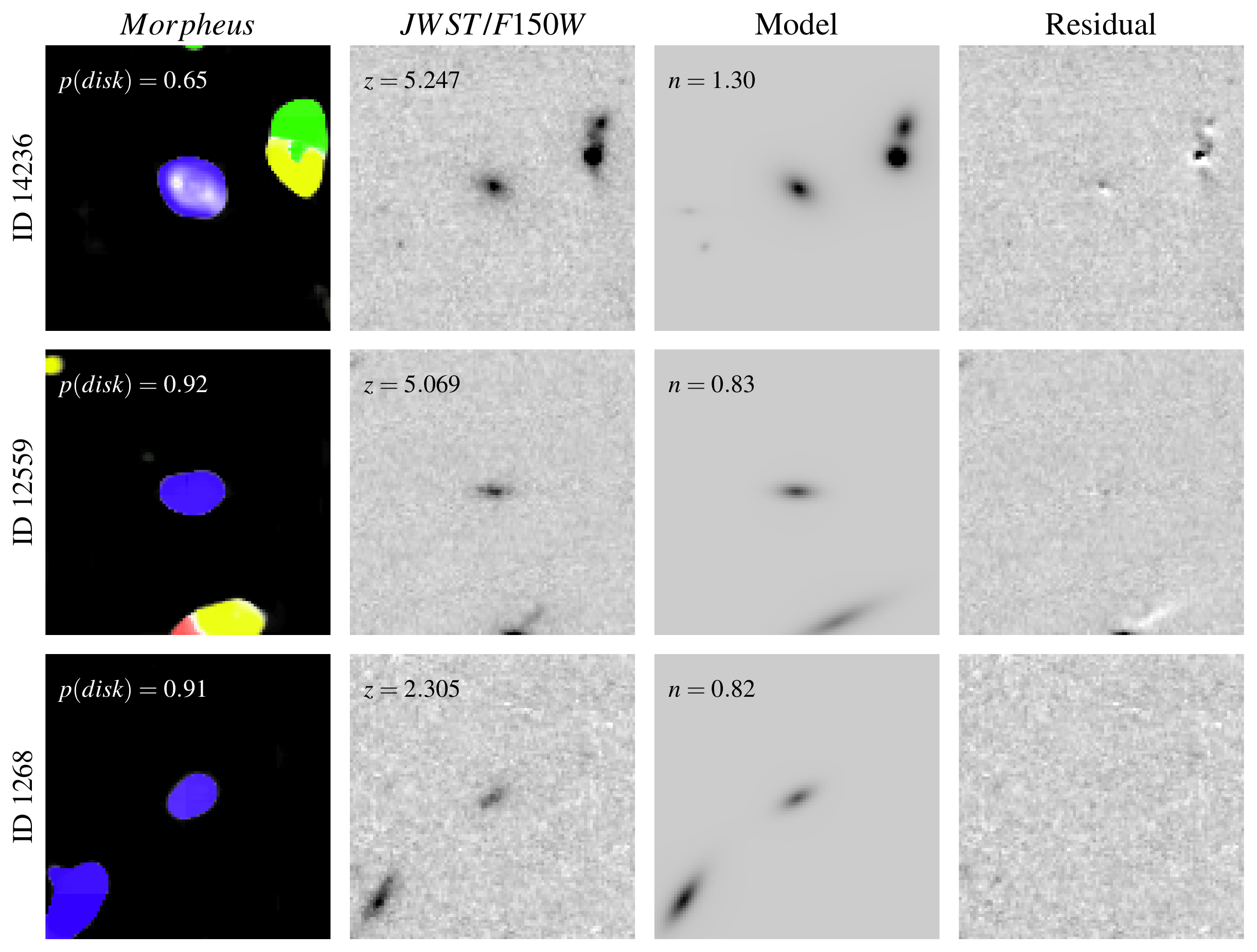}
\caption{High-redshift disk galaxy candidate
identification and modeling. The CEERS \JWST{} $F150W$
images were analyzed using the AI/ML model \Morpheus{} to produce pixel-level classification images, and objects
were selected as disk candidates based on SNR image-weighted
averages of the pixel-level classifications. The left panels show the \Morpheus{} pixel-level classification of three objects (\disk:blue, \irr:green, \psc:yellow, \spheroid:red, indeterminant:white), with $p(\disk)\approx0.65-0.92$.
These objects were cross-referenced with the CANDELS photometric redshift catalog \citep{stefanon2017a} to select objects at $z>2$ (redshifts noted in center left panels). Surface
brightness models of the
\JWST{} $F115W$, $F150W$, $F200W$ object cutouts 
(3''$\times$3''; $F150W$ shown center left) were modeled
using the \ForcePho{} code (Johnson et al., in prep.), and
\citet{sersic1968a} profile indices determined (center right panels). The small residuals between the \JWST{} $F150W$ images
and models indicate high-quality fits (right panels).
All grayscale images shown are displayed with the same linear stretch.
\label{fig:objects}}
\end{figure*}

Figure \ref{fig:sersic} shows a histogram of the
best-fit S\'ersic indices for the entire
\Morpheus{} high-redshift
disk sample from \ProFit{} modeling. 
The vast majority
(183/190; 96\%) of the successful fits displayed ``disky''
($n<2$) S\'ersic indices, with a mean of $\av{n}=1.04$
and median $\hat{n}=0.93$. The \ForcePho{} model
fits show good agreement, with mean and median
S\'ersic indices of $\av{n}=1.06$
and $\hat{n}=0.85$, and 97\% of the disk candidates
displaying $n<2$.
Figure \ref{fig:objects} presents three example
high-redshift disk galaxy candidates identified
by \Morpheus{} with photometric redshifts of
$z\sim2.3-5.2$.
Shown are the \Morpheus{}
pixel-level classification maps for each object
(left panels), their \JWST $F150W$ flux images
(center left panels), the
best-fit \ForcePho{} models of their surface
brightness distributions (center right panels),
and the residuals between the model and the
data (right panels). As these
examples demonstrate, the objects 
\Morpheus{} classified as disks appear flattened
and have best-fit S\'ersic profiles with indices
$n\sim1$ as expected for exponential disks.
In summary, the AI/ML classifications
from \Morpheus{} and the traditional S\'ersic 
index morphological correspondence show
excellent agreement. See H20 for more details on
how \Morpheus{} classifies galaxies with S\'ersic profiles.

Beyond S\'ersic index, the distribution of
axis ratios can be used to assess the fraction
of our candidates that are likely to be disks
rather than prolate, potentially triaxial spheroids.
Both the \ForcePho{} and \ProFit{} models show
very consistent axis ratios, with typical $\av{b/a}\approx0.4$
with a spread $0.2<b/a<1$. We follow \citet{law2012a}
and perform Monte Carlo simulations of
disks and spheroid projected axis ratio distributions.
We then use a Kolmogorov-Smirnov test to compare
consistency between the model and observed axis ratio
distribution.
Since the distributions are not uniform, and given
prior \HST{}
results \citep[e.g.,][]{law2012a,vanderwel2014b,zhang2019a},
we expect spheroids to contribute to the population of
flattened, low-S\'ersic objects and so not all of our
candidates will truly be disks. 
Given the measured axis ratio distributions, we find
an upper limit (3$\sigma$) that
57\% of the sample can be pure disks with $b/a<0.25$
with the remainder being triaxial or prolate spheroids. We
cannot yet rule out a mix of triaxial spheroids
with a wide range of axis ratios contributing to the
majority of the sample.
We also note that there is
evidence from H20 that \Morpheus{} is incomplete for
face-on disks, and we expect this issue to be exacerbated
for faint objects. Correcting for this incompleteness would
raise the upper limits on pure disks.
We leave a more detailed analysis for future
work, but suggest that kinematic measurements from spectroscopic
observations will be required to confirm our candidates as 
high-redshift disks.

\section{Discussion}
\label{sec:discussion}

The ability of \JWST{} to
perform AI/ML-aided identification
of distant disk galaxy candidates
will add to the growing observational
information we have on how early
disk galaxies might form in our
$\LCDM$ cosmology. 
Many of our disk candidates lie at
$z\sim2-3$ where there are substantial
constraints on disk
kinematical structure \citep[e.g.,][]{forster-schreiber2020a},
but less is known about the high-redshift end of our sample. 
Recent
\emph{Atacama Large Millimeter/Submillimeter Array}
(\ALMA{}) observations, when
combined with stellar population modeling,
suggest that early star-forming
main-sequence galaxies are extremely
gas-rich \citep[$\sim90\%$;][]{heintz2022a}.
\ALMA{} observations of [CII] also show evidence
of rotating disk galaxies as early as $z\sim5.5$
\citep{herrera-camus2022a} and rotation in 
quasars as early as $z\sim6.5$ \citep{yue2021a}.
As techniques for identifying high-redshift
disk galaxies with \JWST{} become more refined,
including the first results from
AI/ML techniques presented here, possible targets
for gas kinematical measurements will become
more prevalent. In principle, the \JWST{} NIRSpec
integral field unit spectrograph could also
help reveal the kinematics of these
disk candidates and confirm or refute the
connection between the visual morphology and
dynamical structure of these objects.

Early forming disk galaxies
place important constraints on our theories
of disk galaxy formation. While 
cosmological simulations of
disk galaxy formation found only moderate
success \citep[e.g.,][]{robertson2004a,governato2007a},
more recent simulations 
have made substantial advances in 
reproducing disk galaxy
morphologies \citep[e.g.,][]{aumer2013a,marinacci2014a,wetzel2016a,grand2017a,el-badry2018a,pillepich2019a,libeskind2020a}.
However, disk formation at early epochs
still remains challenging theoretically,
especially for systems with large rotational
support. 
The large $V/\sigma$ seen
in some early-forming systems
\citep[e.g.,][]{neeleman2020a,rizzo2020a,rizzo2021a,fraternali2021a,lelli2021a,tsukui2021a}
does not appear in the bulk population of $z\sim2-5$ disks simulated cosmologically \citep[e.g.,][]{pillepich2019a},
although there are recent reports of success for individual simulated
galaxies
\citep{kretschmer2022a,gurvich2022a,segovia_otero2022a} and very early populations of disks at $z>8$ \citep{feng2015a}.
The presence of these early disks may be sensitive to the galaxy
mass \citep{gurvich2022a} and the gas-rich merger history \citep[e.g.,][]{robertson2006a}.
Identifying candidate early disk galaxies and characterizing their
population properties will help us refine our picture for
disk galaxy formation and understand the cosmological evolution of
galaxy morphology. The combination of \JWST{} and AI/ML techniques
like \Morpheus{} appears promising for identifying distant
disk galaxy candidates for further study.

\section{Summary}
\label{sec:summary}

We present first results
applying AI/ML analysis
to \JWST{} imagery, using
the
\Morpheus{}
deep-learning
framework for pixel-level
astronomical image analysis
with the \JWST{}
Cosmic Evolution Early Release
Science Survey (CEERS; Finkelstein et al., in prep.) data
in the Extended Groth Strip (EGS)
to identify high-redshift ($z>2$)
disk galaxy candidates.
We selected objects with dominant
\disk{}
classifications determined
by \Morpheus{} ($p(\disk)>0.5$) and photometric
redshifts $z>2$ as determined
by the CANDELS survey 
catalogs \citep{stefanon2017a},
identifying 202 high-redshift
disk galaxy candidates.
We then use the \ProFit{}
surface brightness distribution fitting
code \citep{robotham2017a} to
infer single component \citep{sersic1968a}
profiles from the \JWST{} $F150W$ images.
We find the high-redshift disk galaxy
candidates identified by \Morpheus{}
predominantly display exponential (S\'ersic
$n=1$) profiles common for disk galaxy
morphologies, with an average S\'ersic
index of $\av{n}=1.04$. We conclude that
\Morpheus{} identifies galaxies
as disks that would be so classified
by traditional methods, and that \Morpheus{}
shows surprising efficacy at disk classification
even as it was trained on \HST{} $F160W$ images
and was applied to the \JWST{} $F150W$ mosaics
without any modification.
The identification
of high-redshift disk galaxy candidates
in \JWST{} imagery suggests that disk
galaxies may form early, with candidates
in our sample displaying photometric redshifts
as early as $z\sim5$, and may indicate an
early epoch of gas-rich disk formation through
hierarchical processes in the early universe
\citep[e.g.,][]{robertson2006a}. This work motivates
further spectroscopic observations of 
high-redshift disk candidates to constrain
their kinematical structure.

\begin{acknowledgments}
BER thanks Alice Shapley for a helpful discussion that in part motivated
this work.
BDJ, BER, CNAW, DJE, and MJR acknowledge support from the NIRCam Science Team
contract to the University of Arizona, NAS5-02015.
The authors acknowledge use of the lux supercomputer at UC Santa Cruz,
funded by NSF MRI grant AST 1828315.
Funding for this research was provided by the Johns Hopkins University, Institute for Data Intensive Engineering and Science (IDIES).
AJB acknowledges funding from the “FirstGalaxies” Advanced Grant from the European Research Council (ERC) under the European Union’s Horizon 2020 research and innovation programme (Grant agreement No. 789056).
JW gratefully acknowledges support from the ERC Advanced Grant 695671, ``QUENCH'', and the Fondation MERAC.

\end{acknowledgments}

\software{astropy \citep{astropy2013a,astropy2018a},  
          \ForcePho{} (Johnson et al., in prep.),
          \Morpheus{} \citep{hausen2020a},
          \ProFit{} \citep{robotham2017a},
          TensorFlow \citep{tensorflow2015a},
          WebbPSF \citep{perrin2014a}
          }



\end{document}